\documentclass[twocolumn,showpacs,preprintnumbers,amsmath,amssymb,superscriptaddress,aps,prl]{revtex4}
\usepackage{color,graphicx}
\usepackage{bm}
\usepackage{SIunits}
\usepackage{microtype}

\begin{document}

\title{Nano-crystalline inclusions as a low-pass filter for thermal transport in a-Si}

\author{T. Damart}
\affiliation{Institut Lumi\`ere Mati\`ere, UMR 5306 Universit\'e Lyon 1-CNRS, Universit\'e de Lyon, F-69622 Villeurbanne, France}
\author{V. M. Giordano}
\affiliation{Institut Lumi\`ere Mati\`ere, UMR 5306 Universit\'e Lyon 1-CNRS, Universit\'e de Lyon, F-69622 Villeurbanne, France}
\affiliation{Univ. Grenoble Alpes, SIMAP, F-38000 Grenoble, France}
\affiliation{CNRS, SIMAP, F-38000 Grenoble, France}
\author{A. Tanguy}
\affiliation{Institut Lumi\`ere Mati\`ere, UMR 5306 Universit\'e Lyon 1-CNRS, Universit\'e de Lyon, F-69622 Villeurbanne, France}

\begin{abstract}
We use atomistic simulations to study the resonant acoustic modes and compare different calculations of the acoustic mean-free path in amorphous systems with nanometric crystalline spherical inclusions. We show that the resonant acoustic properties are not a simple combination of the vibrations in the inclusions and in the amorphous matrix. The presence of the inclusion affects the transport properties mainly in the frequency range separating simple scattering from multiple scattering processes. However, propagation of acoustic wavepackets is spatially heterogeneous and shows that the amorphous/crystalline interface acts as a low energy pass filter slowing down the high kinetic energy motion whatever the vibration frequency. These heterogeneities cannot be catched by the mean free path, but still they must  play an important role in thermal transport, thus raising the question of the correct modeling of thermal transport in composite systems.
\end{abstract}


\maketitle

\section{Introduction}

The understanding of the propagation of vibrational excitations in disordered materials represents still a challenge for modern research.
In dielectric systems these vibrations are responsible for the thermal transport, and thus the low value and the peculiar temperature dependence of the thermal conductivity in glasses~\cite{zeller_pohl_1971}. Although many theories have been developed over the years, a complete understanding of the mechanisms responsible for the different behavior with respect to crystalline materials has not been achieved.
For longtime the “so-called” thermal anomalies (peak in the specific heat and plateau in the thermal conductivity around 10 K) have been ascribed to the existence of glass-specific soft modes which pile up to give raise to a peak in the reduced density of states at energies in the 0.2-1 THz range: the Boson Peak (BP). Recent investigations suggest however that the BP does not arise from new modes but is the counterpart of the first Van Hove singularity of the corresponding crystalline system~\cite{chumakov_role_2014}. 
Still, in the BP energy range vibrational modes appear to be strongly scattered, so that they are envelopes of different modes with a rapidly decreasing lifetime, which could be at the origin of the plateau in the temperature dependence of the thermal conductivity. 
A classification between propagons (low energy plane waves modes), diffusons (reduced-life time modes beyond the BP energy) and locons (localized modes at high energy) has been proposed by Allen and Feldman and reported again quite recently by Parshin and Beltukov~\cite{beltukov_ioffe-regel_2013, allen_diffusons_1999}. 
The understanding of thermal transport in glasses becomes an important issue also in view of the application of disordered materials in devices requiring very low thermal conductivities. Indeed, semi-conductive glasses have been recently investigated for thermoelectric applications~\cite{ctfm_1, ctfm_2, ctfm_3, ctfm_4, ctfm_5, ctfm_6, ctfm_7}, where the lowest possible thermal conductivity is required, associated with a good electric conductivity.  It is however difficult to get these two properties in a single system: for this reason composite materials made of alternating amorphous and crystalline systems can be promising. Indeed, super-lattices built from alternated crystalline and amorphous layers of silicon have been recently proposed as a novel way to deteriorate the thermal conductivity of silicon while keeping intact its electronic conductivity~\cite{lanord2014}.
These applications raise the question of what happens to vibrational modes when they cross the interface between a disordered and an ordered material.
Here we address this question comparing the vibrational modes in amorphous silicon and in a composite material made of a crystalline inclusion embedded in an amorphous matrix. More specifically, we investigate the nature of the vibrational modes in the composite material and show that at high energy they distinctly differ from the modes in the amorphous system. Interestingly, we find that the presence of the interface perturbs a travelling wave-packet slowing down the components with the highest kinetic energies, thus acting as a low-pass filter for thermal transport. 

In this article, we first focus on the vibrational resonant properties of the samples before investigating the effect of the crystalline inclusion on the propagation of mechanical waves. The article is organized as follows: the first section will present the numerical methods and samples. The second section will report the eigenmodes present in our samples through the computation of the vibrational density of states, participation ratio and visualization of the eigenmodes shape. The last part will focus on the comparison between the mean free paths of mechanical waves obtained through different methods (dynamical structure factor versus wave propagation) in our silicon samples. 

\section{Numerical model and technical details}

In order to prepare the samples and compute the dynamical propagation of waves at constant energy we performed classical molecular dynamics simulation. In parallel, we have computed the resonant vibrations by exact diagonalization of the dynamical matrix of the system.

\subsection{Sample Preparation}

We studied an amorphous silicon system and systems composed of an amorphous matrix containing a spherical crystalline inclusion. For this last configuration, several samples were generated. These systems consist of $N\approx11000$ particles contained in cubic boxes of lengths $L_x=L_y=L_z=60~\angstrom$ with periodic boundary condition (PBC). The interactions between particles are described by the Stillinger-Weber (SW) potential~\cite{stillinger_computer_1985} frequently used to describe amorphous silicon.

The method used to prepare the amorphous solid configuration is described in Ref~\cite{fusco_role_2010}: a silicon crystal of 262 144 particles was first melted at a temperature of 3500~K, and cooled down to 10~K at constant volume at a quenching rate of $10^{12}~K/s$. Then, the energy is minimized using a damped dynamics (see description below). A smaller cube, $60~\angstrom$ large containing $\approx11000$ particles is cut in this sample and a hole of the size of the wanted inclusion is carved in it. A pure crystalline sphere is generated through the repetition of a diamond-like primitive cell and is inserted in the spherical hole of the amorphous sample. At this point, the crystalline and amorphous parts are together, but the bounds at the amorphous/crystal boundary are not established, nor are the positions of the particles realistic with respect to the new system. The sample is thus annealed at $100~K$ during $100~ps$, then, the potential energy of the system is minimized to insure the mechanical stability of the system. To realize this relaxation process the MD code is used to perform damped dynamics. Calculating the forces acting on each particles, we compute for each particle the needed direction for minimizing its potential energy. At each simulation step the particles are slowed down using damping forces opposed to their movements. This procedure can be written as a damped velocity Verlet algorithm where the damping parameter, $\gamma$,  is selected to allow the particles to stop in a close potential energy basin. For our relaxation process we chose $\gamma=0.61~THz$. This relaxation process is performed until a potential energy of $10^{-6}~eV.\angstrom^{-1}$ is reached, putting the system in a local minimum of potential energy with no particles close to plastic instabilities.

Several samples were generated, changing the crystalline inclusion size between 0 and 50\% of the total volume. Here we focus mainly on the sample with 30\% crystalline volume fraction, called CinA30.

\subsection{Structural Characterization}

\begin{figure}
\includegraphics[scale=0.7]{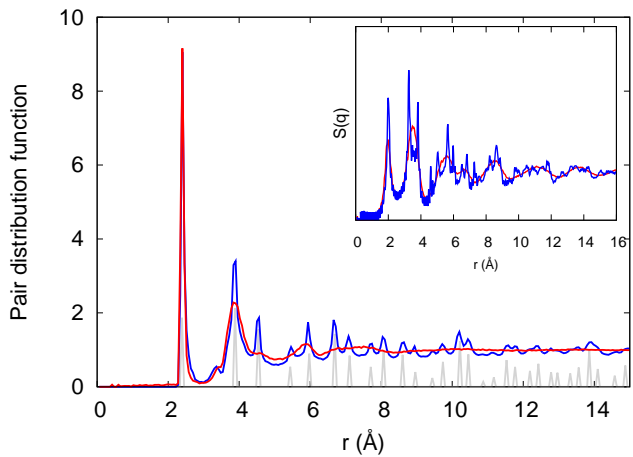}
\caption{Pair distribution function and structure factor (inset) for the amorphous (red line), CinA30 (blue line) samples and crystal (gray). The amplitude of the g(r) of the crystal has been decreased by a factor of 10 to indicate the position of the peaks in a pure crystalline sample.}
  \protect\label{gofr}
\end{figure}

The investigated samples are listed in Table~\ref{modulus} with their shear modulus, bulk modulus, density and sound speeds.
We characterized their structure by computing the pair distribution function (PDF), as reported for the amorphous and CinA30 sample in Fig.~\ref{gofr}. For the amorphous sample, the distribution is characteristic of disordered materials with a local order shown by the sharp first neighbor peak followed by less ordered shells of neighbors. For both the amorphous and CinA30 sample, the nearest neighbor shell is maximum at $2.4~\angstrom$ and ends at $2.9~\angstrom$ in agreement with experimental results on amorphous silicon~\cite{laaziri_high-energy_1999}. For CinA30, peaks at longer range can be seen, which can be related to the presence of the crystalline inclusion where periodicity imposes constant distance between shells of neighbors. The first and second peaks are at the same position with and without inclusion. Indeed, even though the first sample is amorphous, the SW potential imposes the first neighbor distance and the very local organization. For both samples, the pair distribution function shows a shoulder at $3~\angstrom$. This shoulder appears when five fold coordinated atoms are present in the system~\cite{fusco_role_2010}: this over-coordination seems to change the local order and shifts some particles toward this preferential position. It is interesting to notice that the peaks of the crystal in the CinA30 are at the same place that the ones of the pure crystal sample. In the inset of Fig.~\ref{gofr}, we report the structure factor computed from the PDF. The maximum of the structure factor fixes the size of the pseudo Brillouin zone~\cite{giordano_2010}, which, for the amorphous sample, is given by the second peak at $q=3.8~\angstrom^{-1}$.

\begin{table*}
\begin{center}
\begin{tabular}{ccccccc}
\hline
\hline
 & Amorphous & CinA10 & CinA20 & CinA30 & CinA40 & CinA50\\
\hline
Crist. Vol. fraction (\%) & 0 & 10 & 20 & 30 & 40 & 50 \\ 
Inclusion's radius ($\angstrom$) & 0 & 17.27 & 21.77 & 24.91 & 27.42 & 29.54\\
Shear Modulus ($GPa$) & 29.0 & 29.4 & 29.5 & 28.3 & 29.9 & 35.7\\
Bulk Modulus ($GPa$) & 103.0 & 105.0 & 105.7 & 105.8 & 106.6 & 108.1\\
Density ($g/cm^3$) & 2.31 & 2.31 & 2.31 & 2.31 & 2.31 & 2.33\\
Longitudinal sound speed ($m/s$) & 7831 & 7901 & 7924 & 7883 & 7963 & 8175\\
Transverse sound speed ($m/s$) & 3543 & 3568 & 3574 & 3500 & 3598 & 3914\\
\hline
\end{tabular}
\caption{Comparison of the structural properties of the amorphous sample and amorphous samples containing a crystalline volume fraction X from 0 to 50\%, referred to as CinAX.}
\label{modulus}
\end{center}
\end{table*}

\section{Resonant Vibrational Properties}

In order to understand the effects of the inclusion on the vibrational response we studied the eigenmodes of the systems.
We started by computing the dynamical matrix (Hessian matrix) which, if anharmonic interactions between particles are neglected, can be defined as~\cite{dove_introduction_1993, ashcroft_solid_1976}: \begin{equation}M_{ij}^{\alpha\beta}=\frac{1}{\sqrt{m_im_j}}\frac{\partial^{2}E_{pot}}{\partial x_{i}^{\alpha}\partial x_{j}^{\beta}}\end{equation}
$M_{ij}^{\alpha\beta}$ is thus the component of the dynamical matrix that links the displacement of the particle $j$ in the direction $\beta$ to the displacement of the particle $i$ in the direction $\alpha$ through the potential energy variation induced by this displacement. 
If we assume the displacements in the system to be waves of the form \begin{equation}u_{i}^{\alpha}=r_{i}^{\alpha}e^{i\omega t}\end{equation} Newton's law becomes \begin{equation}\sqrt{m_{i}}\omega^{2}r_{i}^{\alpha}\simeq\sum_{j,\beta}M_{ij}^{\alpha\beta}r_{j}^{\beta}\sqrt{m_j}\end{equation}
$\sqrt{m_i}r_{i}^{\alpha}$ and $\omega^{2}$ are the eigenvectors and the eigenvalues of the dynamical matrix while $\omega/2\pi$ are the eigenfrequencies of the system. Numerically, the dynamical matrix is computed in a discrete way, by taking the particles one by one, moving them $0.001~\angstrom$ from their equilibrium position and looking at the change of forces created on all the other particles, for all the directions. Then the exact diagonalization of the dynamical matrix is done using the FEAST Eigenvalue Solver~\cite{_feast_????}, thus obtaining the $3N$ eigenmodes of vibrations and eigenfrequencies of the system.
For the computation of the quantities presented in the following (dynamical structure factor and mean free path), the interactions are described by the dynamical matrix, which is generated using the harmonic part of the SW potential.

\subsection{Density of Vibrational Modes}

\begin{figure}
\includegraphics[scale=0.7]{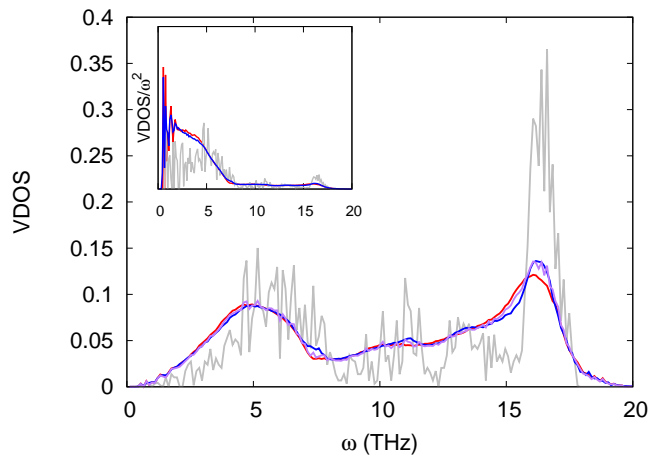}
\caption{Vibrational density of states for the crystal (gray line), amorphous (red line) and CinA30 (blue line) samples. The purple line is the linear combination of the crystal and amorphous VDOS (See text for details). The inset represents $VDOS/\omega^2$ and highlights the boson peak.}
  \protect\label{vdos}
\end{figure}

The vibrational density of states (VDOS) can be obtained by summing the number of eigenfrequencies existing for each frequency interval $\delta\omega=0.11~THz$. The result can be seen in Fig.~\ref{vdos} for the amorphous, CinA30 and crystalline sample. The difference of the VDOS from the Debye's model is represented as $VDOS/\omega^2$ in the inset. An excess of frequency appears from 2~THz to 6~THz, corresponding to the so-called boson peak.

\begin{figure}
\includegraphics[scale=0.7]{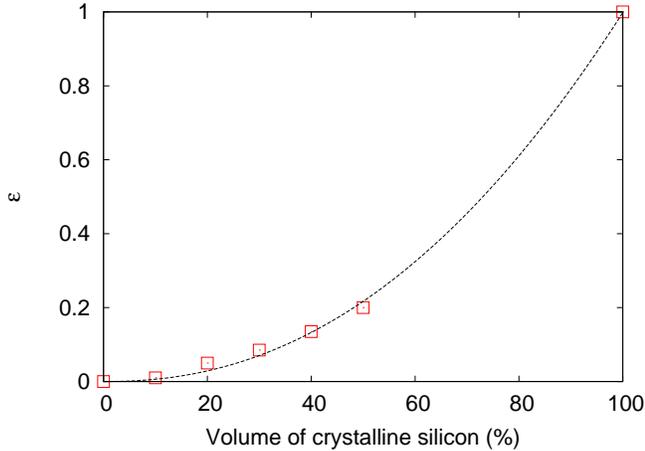}
\caption{Weight $\epsilon$ of the crystalline VDOS in a linear combination for matching the VDOS of systems containing from 0\% to 100\% of crystalline inclusion. The line is the fit with a power law: $f(x)=0.0038x^{2.2}$.}
  \protect\label{dossum}
\end{figure}

\begin{figure}
\includegraphics[scale=0.7]{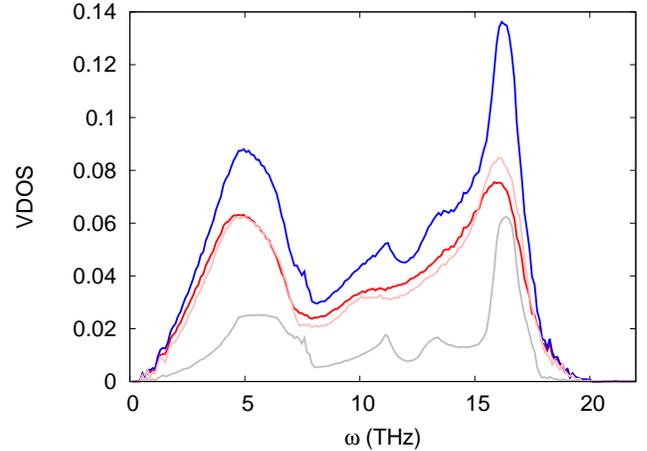}
\caption{Vibrational density of states of the CinA30 (blue line) sample, and partial VDOS computed from the displacement of the atoms of the amorphous (red line) or crystalline (gray line) part of the sample. The pink line represents 70\% of the VDOS of the amorphous sample.}
  \protect\label{dossep}
\end{figure}

The VDOS shows a similar shape for the amorphous and CinA30 samples. The influence of the crystalline inclusion can be seen on the CinA30 VDOS mainly at $7.5~THz$ and from $11~THz$ to $17~THz$. We plot in light blue the linear combination of the pure crystal and the pure amorphous VDOS meant to reproduce the CinA30 VDOS, as done on nanocrystals by Stankov~\textit{et al.}~\cite{stankov_vibrational_2010}. This linear combination is defined as $VDOS_{linear}=\epsilon*VDOS_c+(1-\epsilon)*VDOS_{\alpha}$. The values found for $\epsilon$ in order for the $VDOS_{linear}$ to match at best the $VDOS_{CinA30}$ are much smaller than the crystalline volumic fraction (Fig.~\ref{dossum}). It can also be noticed that even if the $VDOS_{linear}$ and $VDOS_{CinA30}$ match well, some discrepancies exist, in particular at frequencies where the crystal and the amorphous VDOS are very different. These two observations show that the VDOS of the modes of a sample with a crystalline inclusion is not a simple combination of the modes of the two independent crystalline and amorphous systems.
To go further,  we computed the partial VDOS for the amorphous and crystalline part of the CinA30 sample as:
\begin{equation}g_{am}(\omega)=\sum_{i=1}^{3N}\sum_{j\in am}^{N_{am}}\| \vec{u}_j^i\|^2\delta(\omega-\omega_i)\end{equation}
\begin{equation}g_{cr}(\omega)=\sum_{i=}^{3N}\sum_{j\in cr}^{N_{cr}}\| \vec{u}_j^i\|^2\delta(\omega-\omega_i)\end{equation}
Fig.~\ref{dossep} shows that the partial VDOS of the amorphous part is quite similar to the VDOS of the amorphous sample, reported here scaled for a better visibility, thus the crystalline inclusion does not have an influence on the modes of the amorphous part. On the other hand, the VDOS of the crystalline part is a very smoothed version of the one of the pure crystal. It is clear from this figure that the dips in the total VDOS are due to the crystalline part.
This result raises the question of the spatial extension of the eigenmodes and their possible localization at the interface. In order to investigate this point, we analyze more in details the eigenmodes in the following section.

\subsection{Vibrational Eigenmodes}

\begin{figure}
\includegraphics[scale=0.7]{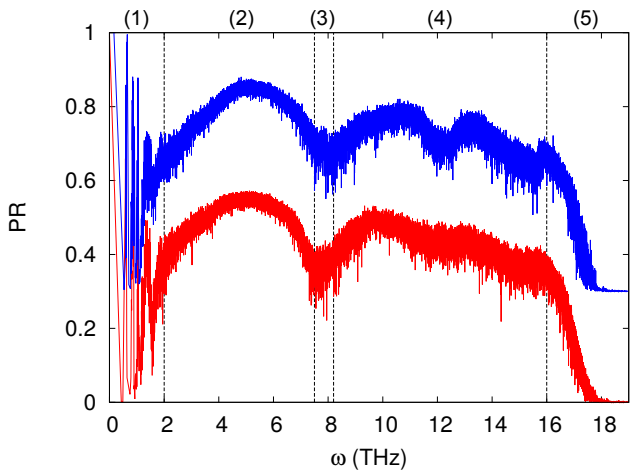}
\caption{Participation ratio (PR) for the amorphous (red line) and CinA30 (blue line) sample. The CinA30 PR has been shifted along the y-axis for the sake of visibility. The dashed lines separate the five frequency regimes discussed in the text.}
  \protect\label{participation}
\end{figure}

In order to quantify the spatial extent of the modes we compute the participation ratio (PR), which expresses the fraction of particles in the system taking part to the motion for a given mode. It is defined as: \begin{equation}PR=\frac{1}{N} \frac{(\sum_{i=1}^N\|\vec{u_i}\|^2)^2}{\sum_{i=1}^N\|\vec{u_i}\|^4}\end{equation} with $\vec{u_i}$ the displacement of the ith particle and N the number of particles in the system. The results obtained for the amorphous and CinA30 samples are shown in Fig.~\ref{participation}.
The general shape of the PR follows previously obtained results for amorphous silicon~\cite{allen_diffusons_1999, bodapati_vibrations_2006}. Among other differences, the PR of the CinA30 sample is slightly shifted toward higher frequencies, reflecting the higher wave's velocity found for this sample (See Table~\ref{modulus}). 

\begin{figure*}
  \centering
  \includegraphics[width=.16\linewidth]{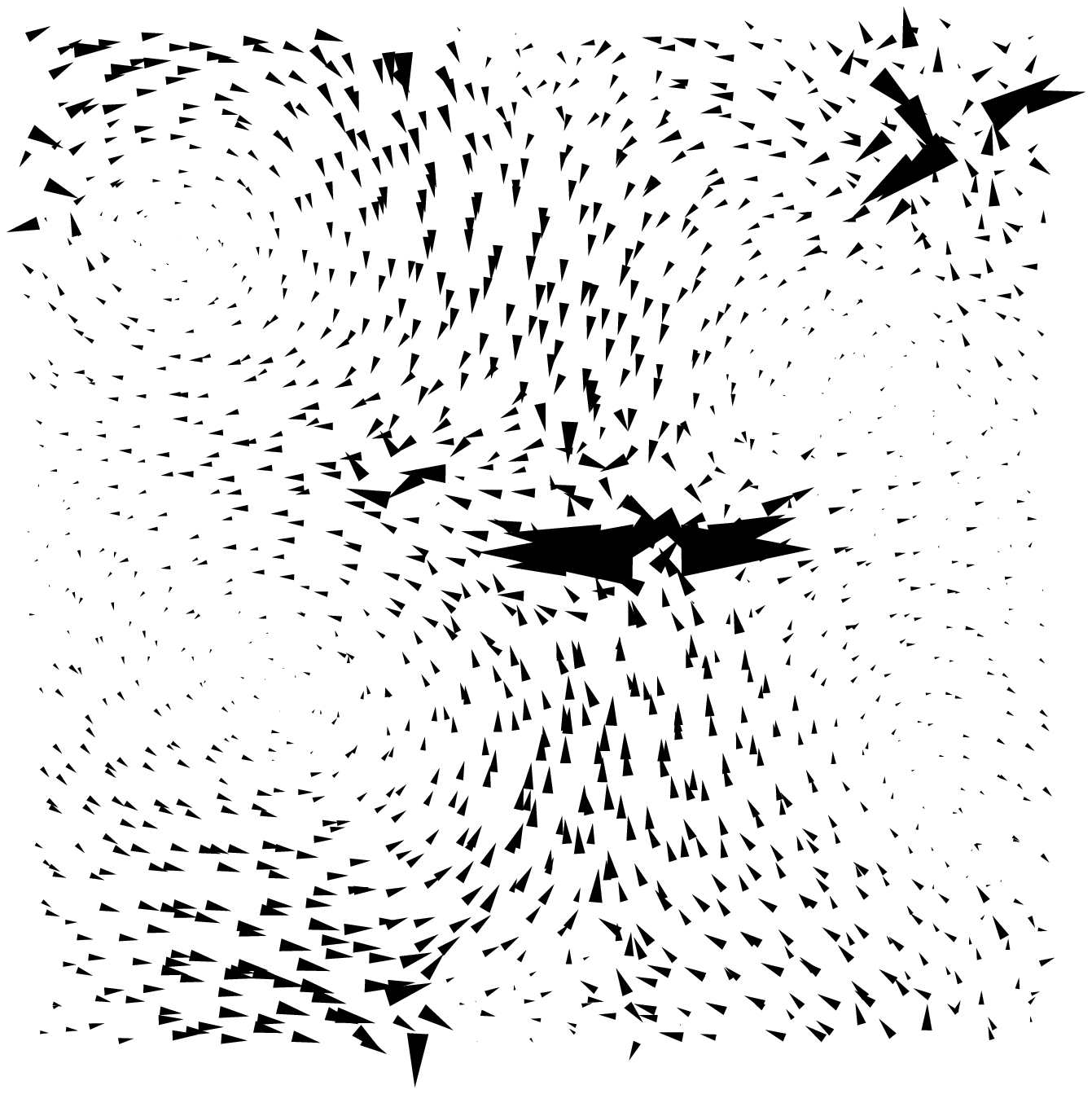} \qquad
  \includegraphics[width=.16\linewidth]{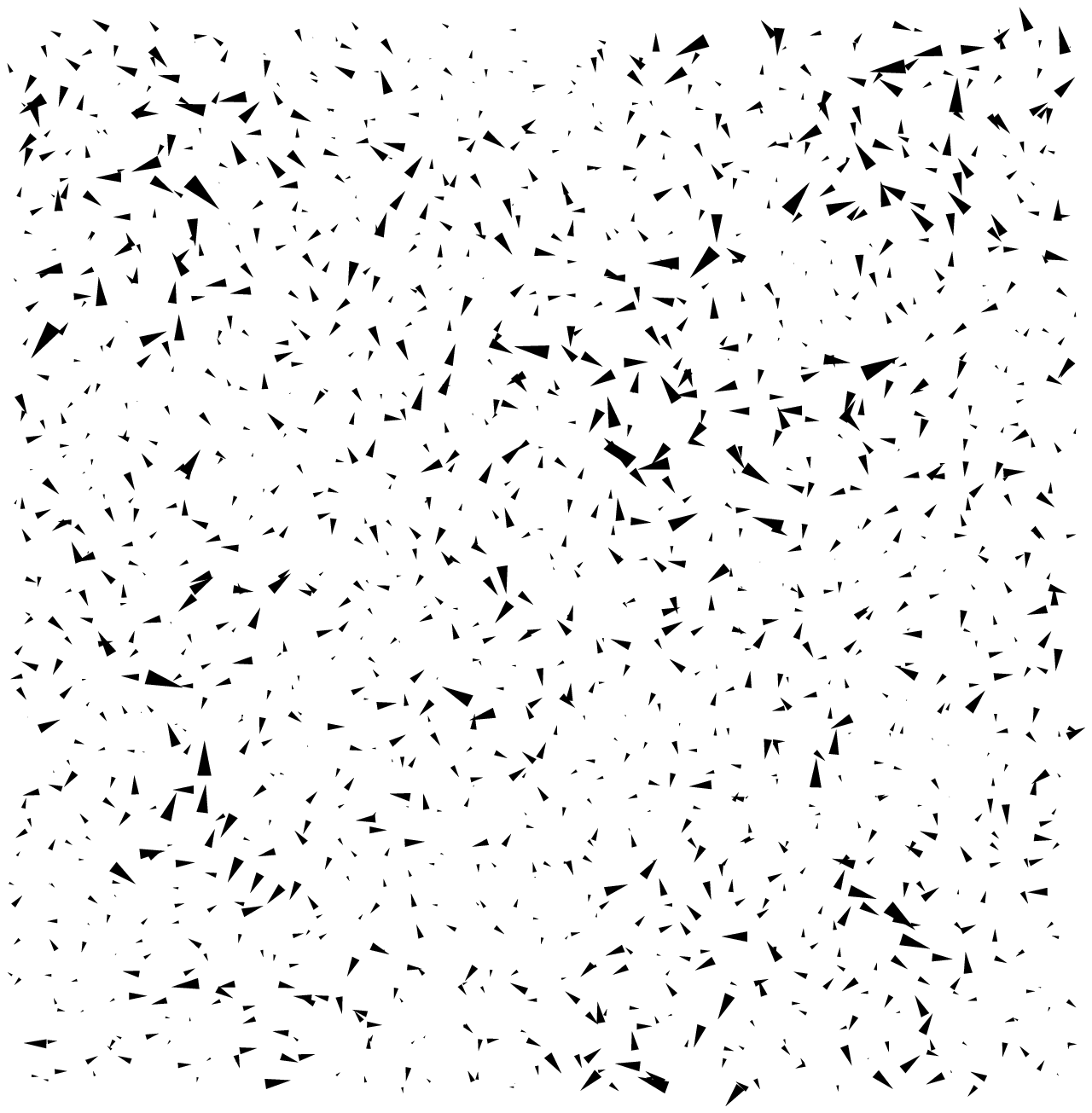} \qquad
  \includegraphics[width=.16\linewidth]{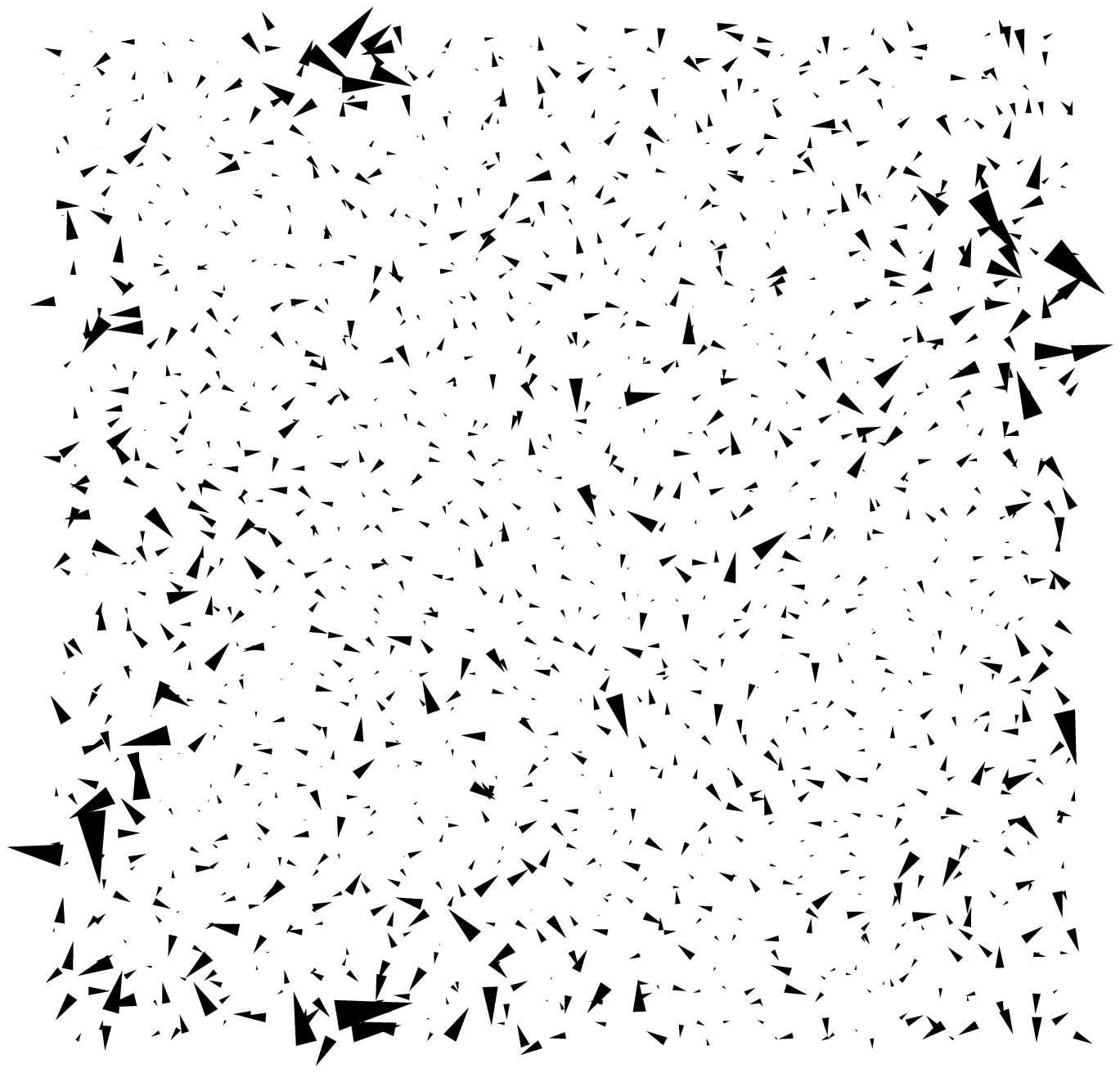} \qquad
  \includegraphics[width=.16\linewidth]{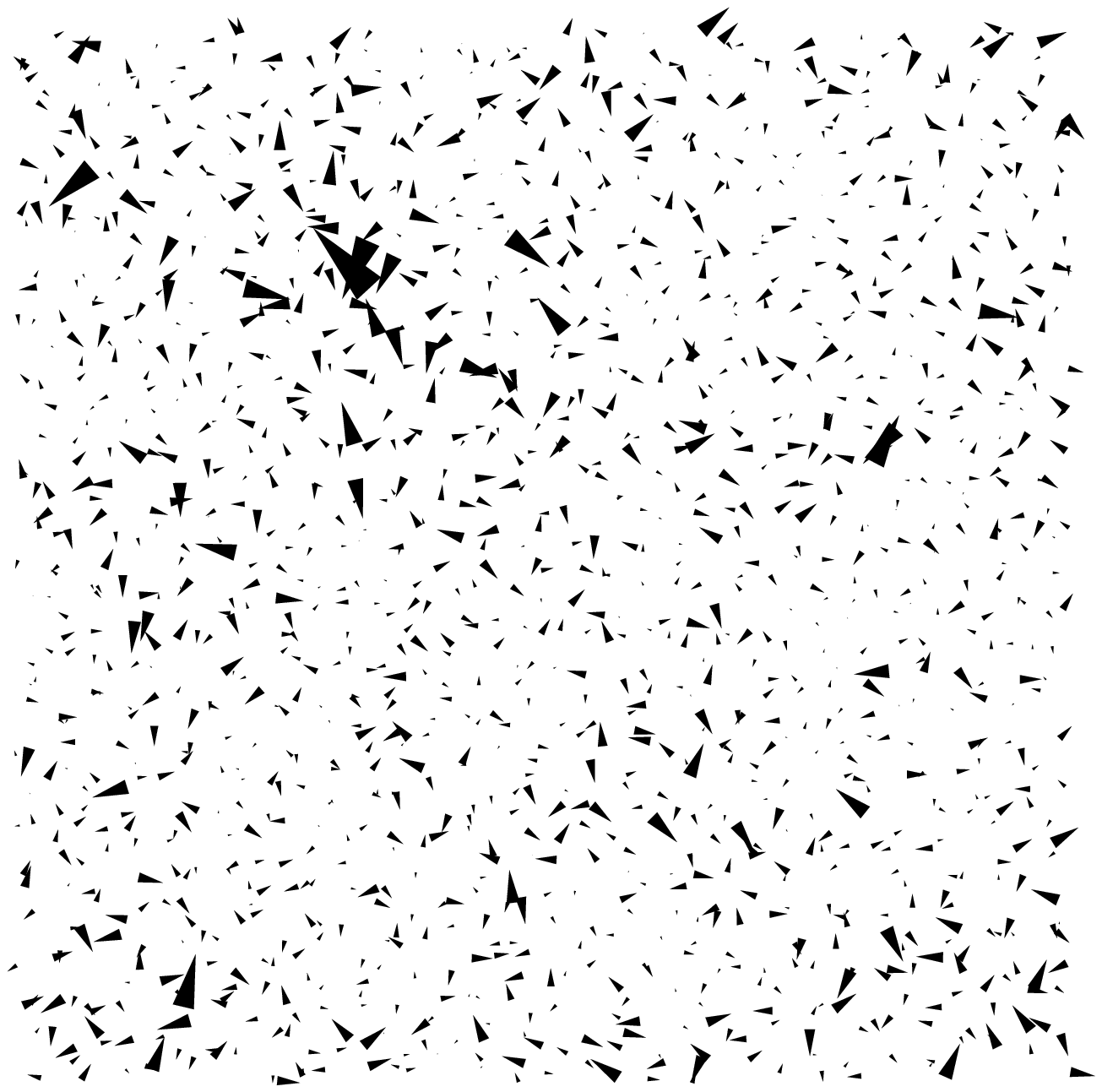} \qquad
  \includegraphics[width=.16\linewidth]{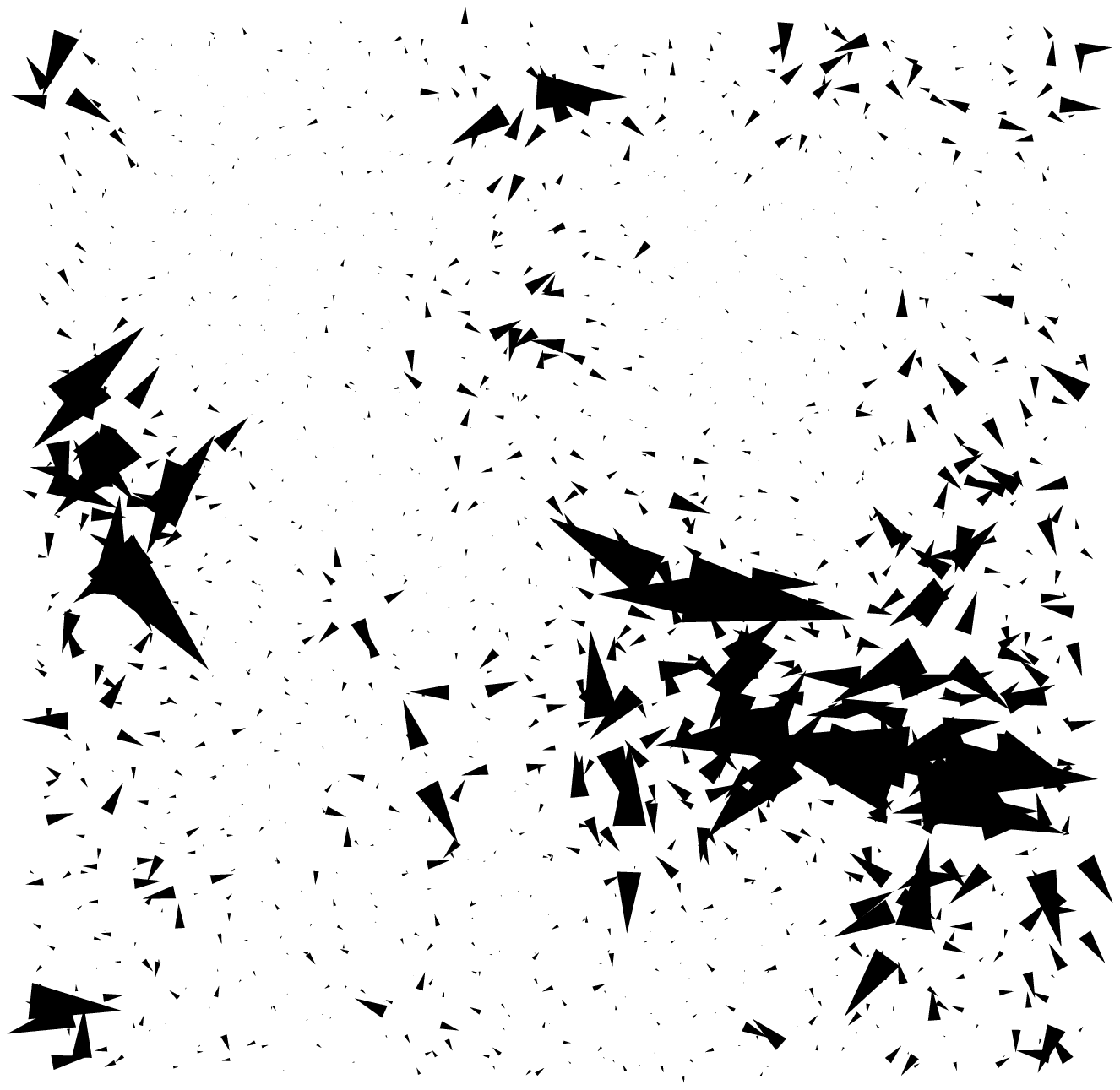}
  \caption{Representation of different modes of vibration for the amorphous sample. The frequencies of the modes are from left to right 0.795 (Regime 1), 4.817 (Regime 2), 7.663 (Regime 3), 12.107 (Regime 4) and 17.451~THz (Regime 5).}\protect\label{modesA}
\end{figure*}

\begin{figure*}
  \centering
  \includegraphics[width=.16\linewidth]{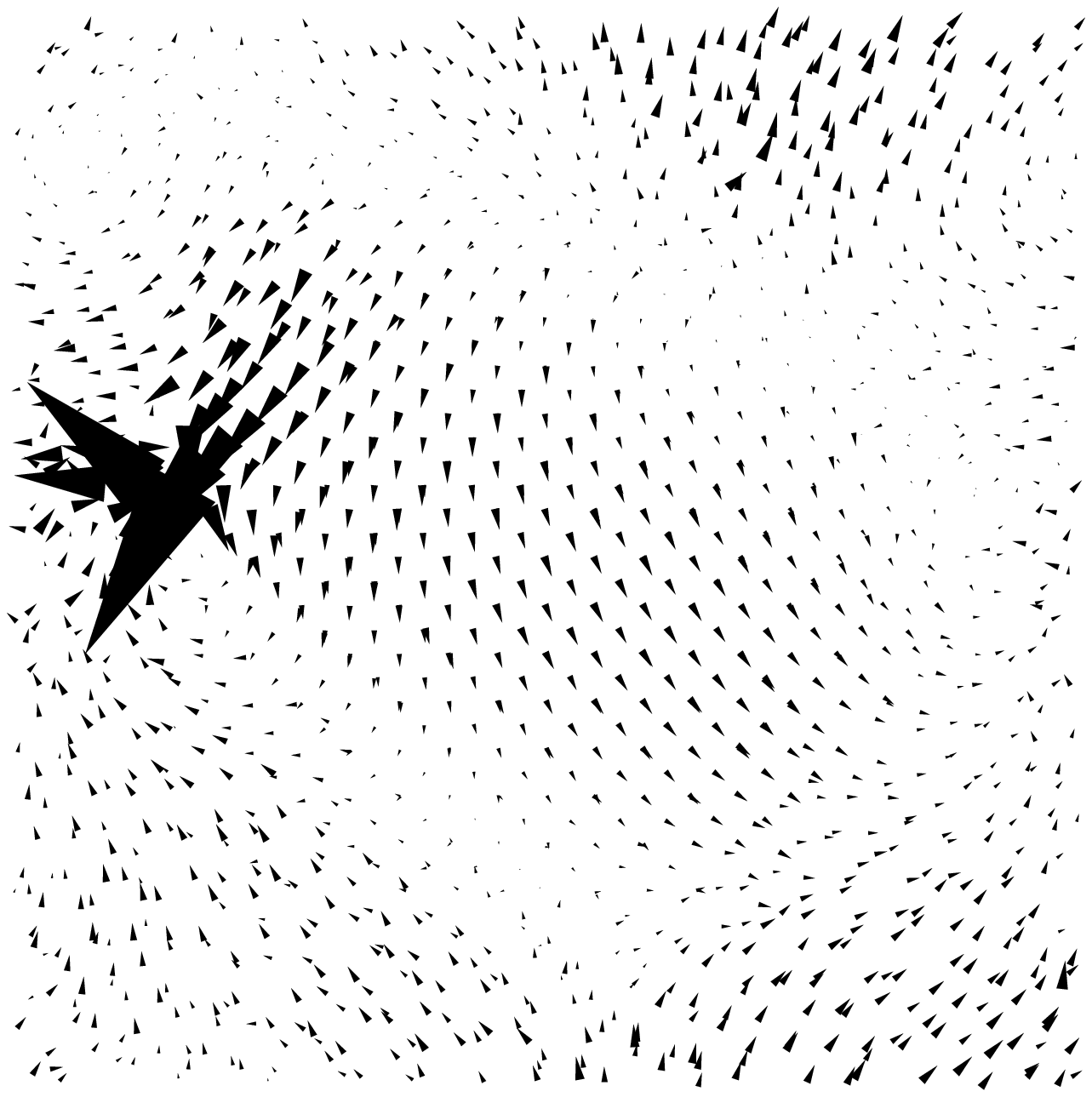} \qquad
  \includegraphics[width=.16\linewidth]{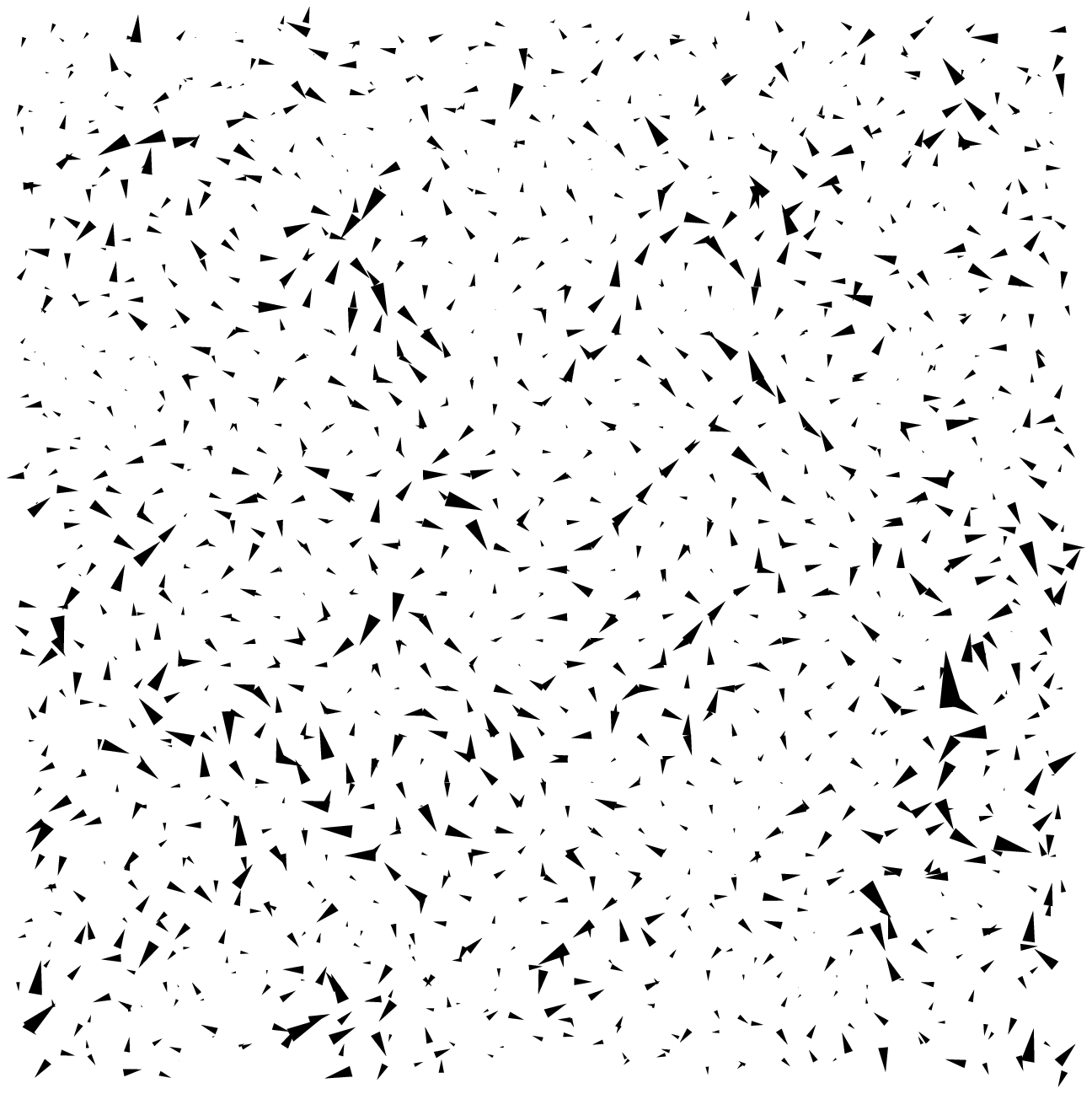} \qquad
  \includegraphics[width=.16\linewidth]{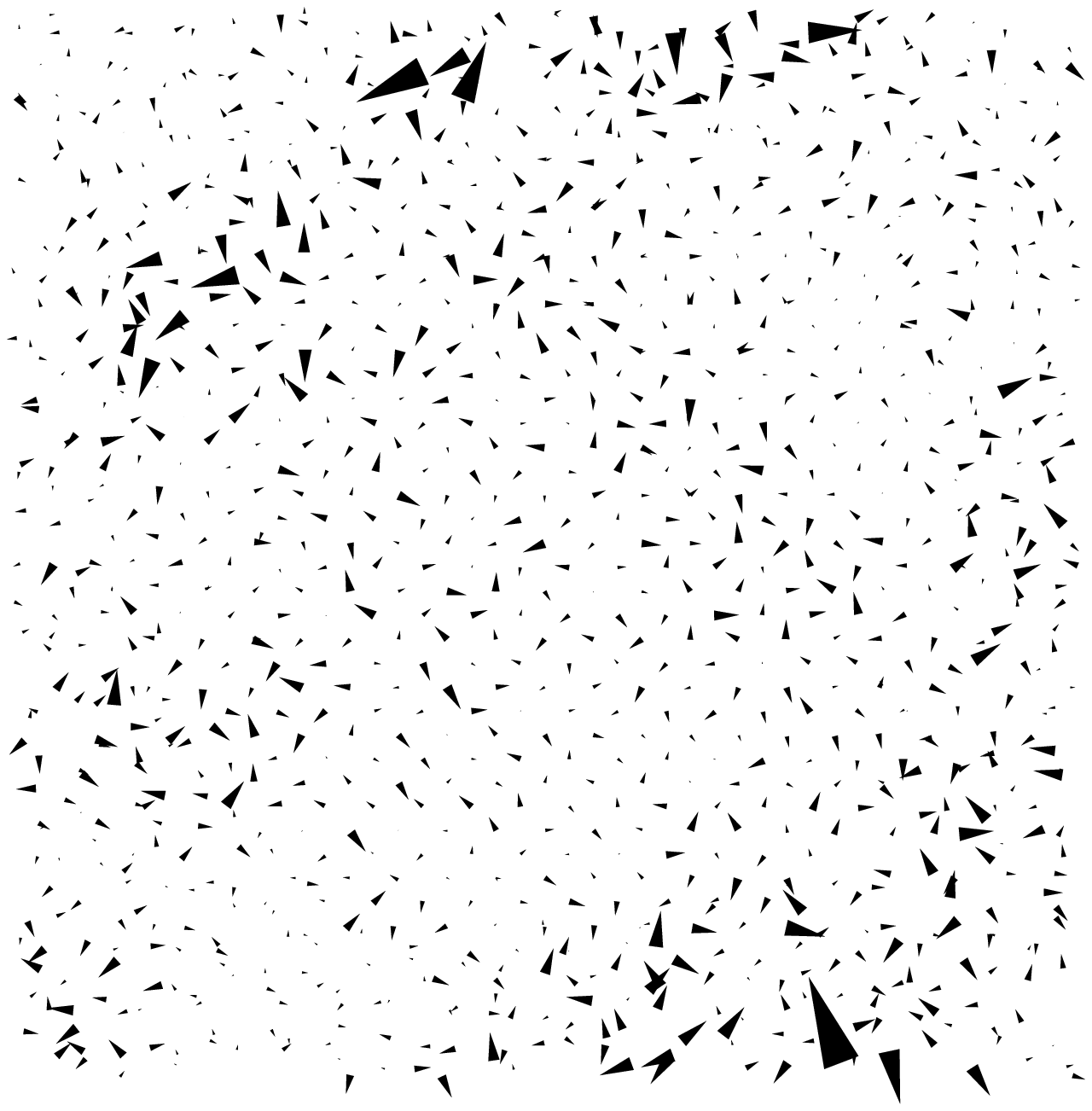} \qquad
  \includegraphics[width=.16\linewidth]{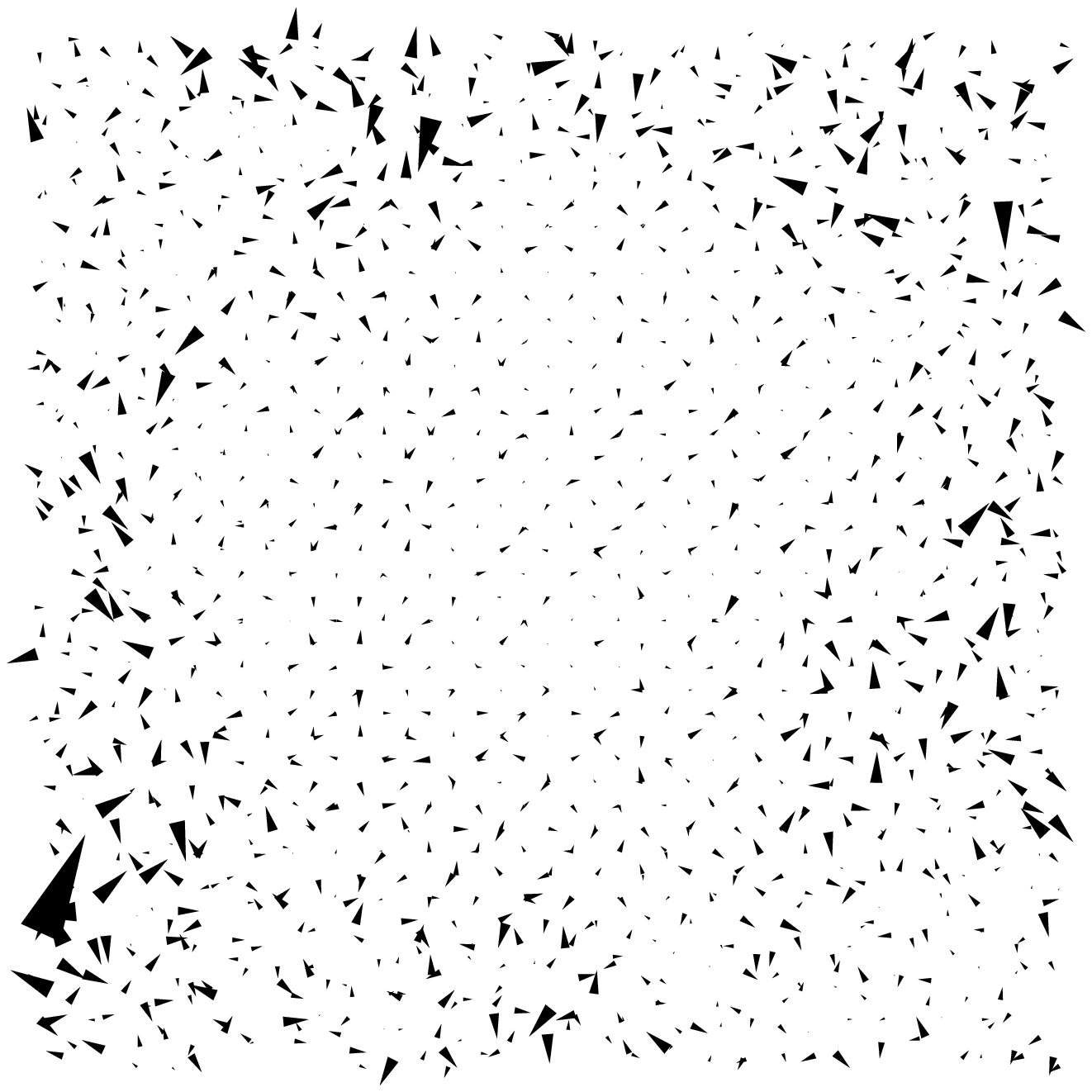} \qquad
  \includegraphics[width=.16\linewidth]{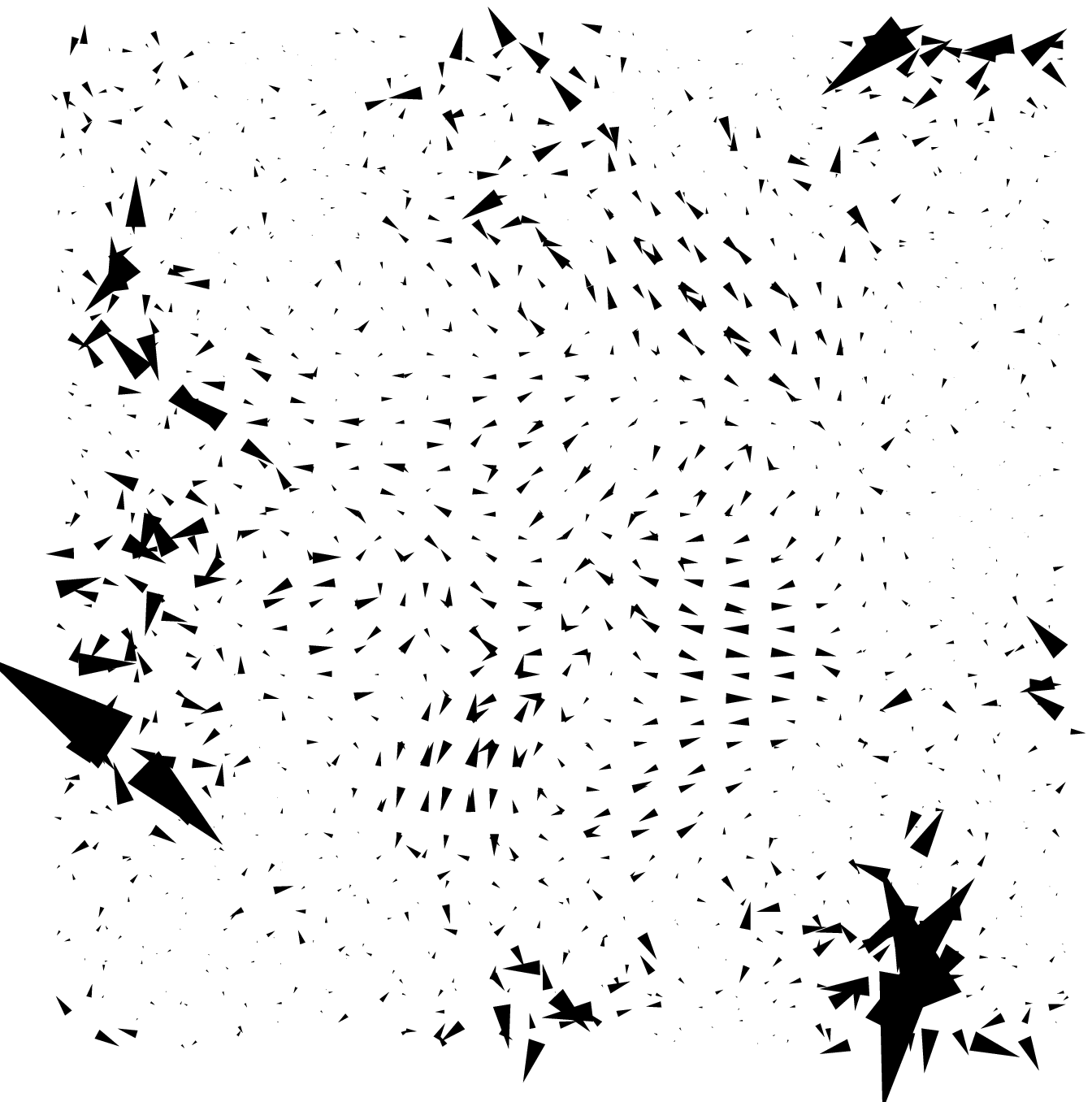}
  \caption{Representation of different modes of vibration for the CinA30 sample. The frequencies of the modes are from left to right 0.816 (Regime 1), 4.903 (Regime 2), 7.803 (Regime 3), 12.205 (Regime 4) and 17.511~THz (Regime 5).}\protect\label{modesCinA}
\end{figure*}

Five different regimes of vibration can be identified from the PR (See Fig.~\ref{participation}): we report vibrational modes for these five frequency ranges in Fig.~\ref{modesA} and~\ref{modesCinA} for the amorphous and CinA30 sample respectively. We show on these figures the displacement of the particles in a $8~\angstrom$ thick layer of the sample containing the displacement of highest amplitude. The arrows direction and size represent the direction and amplitude of the particles displacement.

Regime (1): At zero frequency, PR=1 for the three translational modes corresponding to the global translation of the system in the three directions of space. Up to 2.0~THz, there is a mix of high PR (delocalized) modes and low PR modes that can be identified as soft modes~\cite{tanguy_vibrational_2010}. Plane waves are delocalized over the whole system leading to the high but wavelength dependent PR seen at these frequencies. Soft modes are the superposition of delocalized modes of long wavelength, and displacements involving only small selected groups of particles. These modes are known to be the markers of imminent plastic deformation (their frequency tends to zero when the system approaches a plastic deformation)~\cite{tanguy_vibrational_2010} and it is known that their number is much larger in a silicon-like than in a Lennard-Jones system~\cite{tanguy_vibrational_2010}. It is interesting to notice that the influence of the inclusion is not limited only to the interface. 

Regime (2): This range of frequency shows a broad area of high PR that reaches 55\%. Very characteristic whirlpool-like structures can be observed here (Fig.~\ref{modesA} and Fig.~\ref{modesCinA}), which arise from the disordered arrangement of the particles in the amorphous matrix and have already been studied by one of us~\cite{tanguy_continuum_2002, tanguy_vibrational_2010}. For the CinA30 sample, the crystalline and the amorphous matrix move together. The modes extend over the whole sample leading to a high PR. In this frequency range, the presence of an amorphous/crystal interface does not seem to prevent the propagation of the modes. This is probably related to the fact that both systems have large values of the VDOS in this frequency range. Recent works~\cite{Parshin2014} show that for this range of frequency the modes are almost entirely longitudinal.

Regime (3): Here we observe a drop of participation ratio that reaches a local minimum of 35\% for both the amorphous and CinA30 samples, independent from the presence of the inclusion. The displacements are here localized but less spatially focused than soft modes. 

Regime (4): This region of the PR is characterized by two interesting drops around 12.5 and 15~THz for the CinA30 sample. Precursors of these drops can be seen in the amorphous sample's PR but they become more marked as the size of the inclusion increases. The largest displacements, that show where the modes are localized, are on groups of particles in the amorphous matrix. These modes are free to extend in the purely amorphous sample, where the PR is therefore high. However, in the case of CinA30, the crystalline part stays very still, thus reducing the PR. It could be due to the fact that the crystal has fewer modes of vibration at these frequencies~\ref{dossep}: the modes of vibration cannot extend in the crystalline part where they do not exist. It should be noticed that similar drops of PR have already been observed in amorphous silicon by Allen~\textit{et al.}~\cite{allen_diffusons_1999} and have been interpreted as "resonant modes" that are not localized but temporarily trapped in regions of peculiar coordination. For Allen~\textit{et al.} this interpretation is valid for the drops of PR at low and high frequency. We show here that the four different low PR areas can be explained by different phenomena. More recently, similar drops of PR have been observed in nanocrystalline silicon where their origin was the confinement of these modes on grains interfaces or on grain interiors~\cite{bodapati_vibrations_2006}, as well as in Si/Ge core-shell nanowires where no interpretation was proposed~\cite{hu_significant_2011} and in silicon nanotubes where the low PR modes were localized on the surface of the material~\cite{chen_remarkable_2010}. The original result here is that these modes are localized in the matrix rather than at the interface. Still, one could wonder what would be observed in a crystalline matrix with amorphous inclusion.

Regime (5): For these frequencies, the PR shows a continuous decay to 0\% up to 18~THz, corresponding to modes more and more localized as the frequency increases. As it can be seen in Figg.~\ref{modesA} and~\ref{modesCinA} the localization is now organized in patches containing each several particles while the localized modes observed at lower frequencies were organized around small spots of localization. The high frequency localized modes can be seen as a manifestation of Anderson localization for phonons and have started to be studied in the framework of the multifractal theory applied to participation ratio~\cite{evers_anderson_2008,nicolas_etude_2014}. The shape of these modes in the CAin30 sample is more difficult to interpret because of the presence of optic modes in the crystalline part at these frequencies.

\section{Mean Free Path}

\subsection{Dynamical Structure Factor}

\begin{figure*}
  \centering
  \includegraphics[width=.4\linewidth]{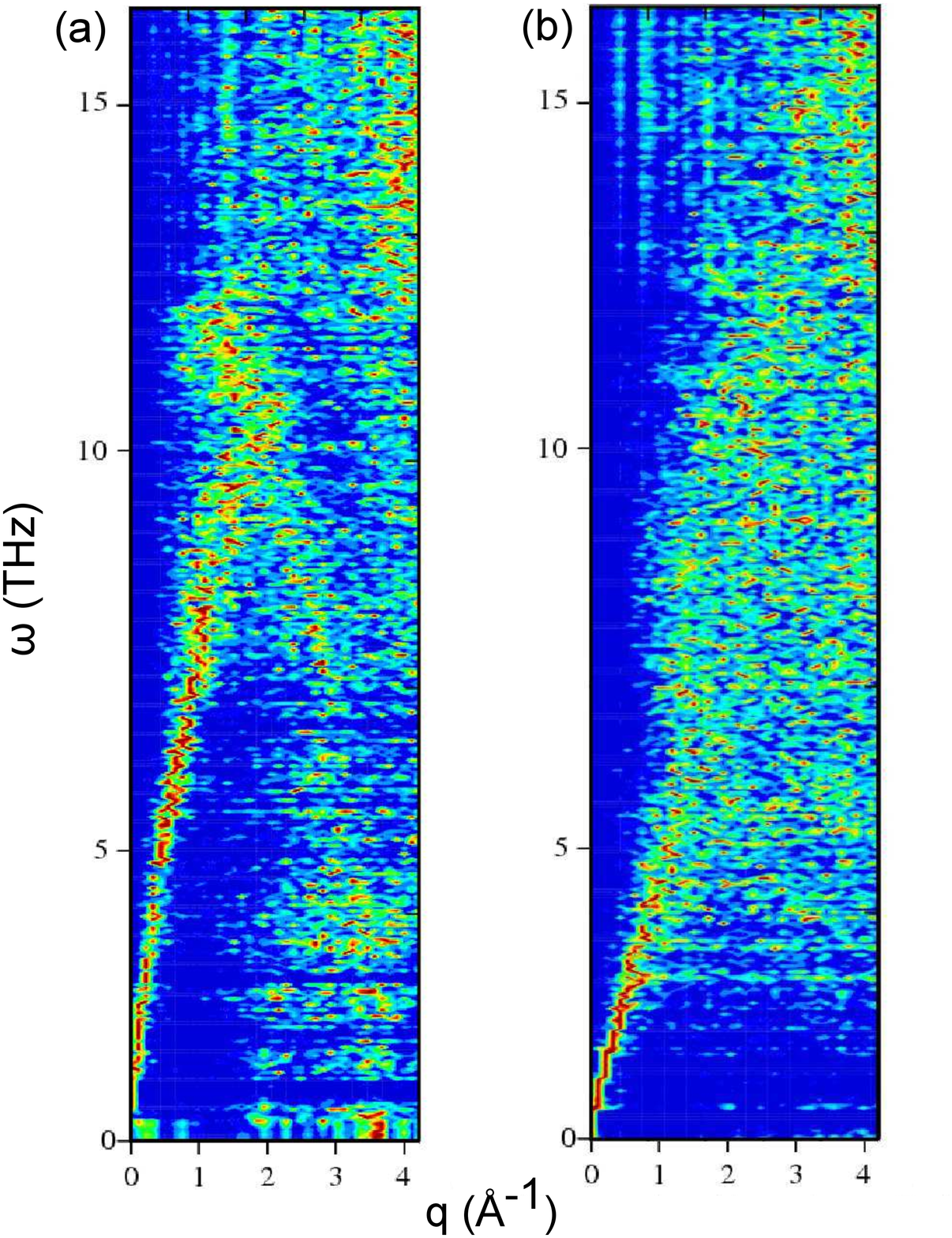} \qquad
  \includegraphics[width=.4\linewidth]{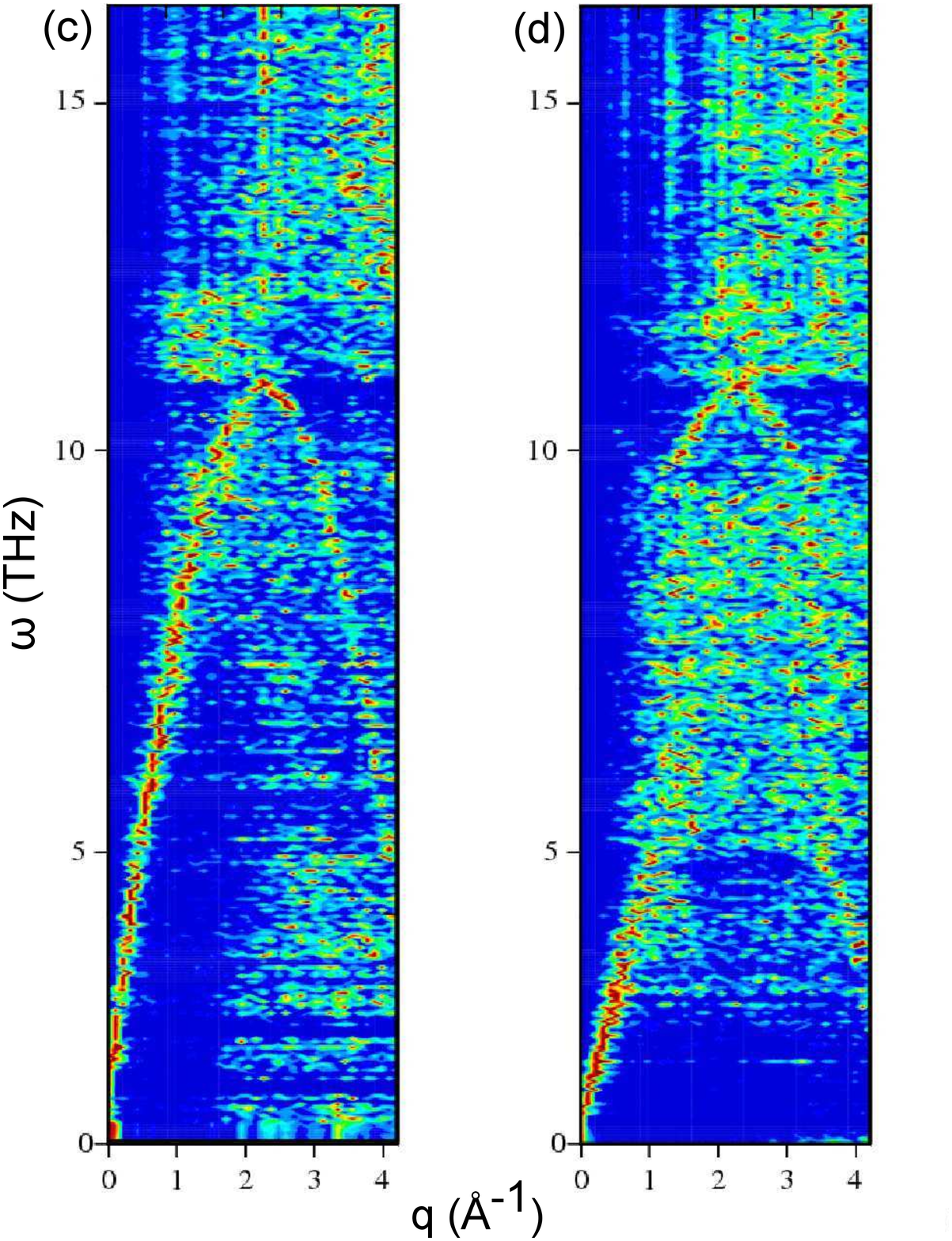} 
  \caption{Dispersion curve for the amorphous sample ((a): longitudinal, (b): transverse) and CinA30 sample ((c): longitudinal, (d): transverse) plotted from the dynamical structure factor. The redness of a point indicates its intensity in the dynamical structure factor normalized such that the most intense point for each fixed frequency is in red.}\protect\label{dispersion}
\end{figure*}

To further study the vibrational properties of our systems, we computed the dynamical structure factor using the method described by Y.~Beltukov~\textit{et al.} in ~\cite{beltukov_ioffe-regel_2013}. At the first time step of the MD run, once the structure is equilibrated, we impose to all atoms speeds randomly generated using a Maxwell-Boltzmann distribution centered at 100.0~K with variance 0.2~K$^2$. The forces between particles are described by the dynamical matrix: only harmonic interactions are taken into account here, and non-linear processes are neglected, thus the MD simulation runs at constant energy (NVE) with a negligeable temperature variation. The output of this run is a file containing the positions of the particles at each time step. Using these data we can compute the dynamical structure factor, $S(\vec{q},\omega)$, for the system by performing a spatial and temporal Fourier transform on the displacement of all the particles for each timestep starting from the beginning of the MD run: \begin{eqnarray}&&S(\vec{q},\omega)=\\&&\frac{2}{NT}\left|\sum_{i=1}^N\int_0^T\vec{u}(\vec{r_i},t)\cdot\vec{m}_q\exp(-i\vec{q}\cdot\vec{r_i})\exp(i\omega t) \, \mathrm dt\right|^2\nonumber\end{eqnarray} where $T$ is the total duration of the MD run and $\vec{m}_q$ a polarization vector parallel or perpendicular to the wavevector $\vec{q}$ to have access separately to the longitudinal or transverse waves. 
Following this procedure, we obtained the longitudinal and transverse dispersion relations (Fig.~\ref{dispersion}). A folding of the acoustic modes can be seen for both samples, centered at $q=1.9~\angstrom^{-1}$ and $q=2.2~\angstrom^{-1}$ for the longitudinal dispersion relation in the amorphous and CinA30 sample respectively. These values are compatible with the pseudo-Brillouin zone border, as determined by the most intense peak of the static structure factor, at $\approx 3.8~\angstrom^{-1}$ (inset of Fig.~\ref{gofr}). Generally speaking the dispersion relation of the CinA30 sample is better defined than in the amorphous one. This is most striking for the transverse waves: for both samples there is a sudden broadening of the dispersion at $\approx3$~THz, suggesting a transition from a propagative to a diffusive regime~\cite{allen_diffusons_1999}. However in CinA30 newly propagative-like modes reappear between 10 and 11~THz, while above only non-dispersive modes are present, compatibly with the fact that only optic modes exist at these frequencies in the crystal.

\begin{figure}
 \includegraphics[scale=0.7]{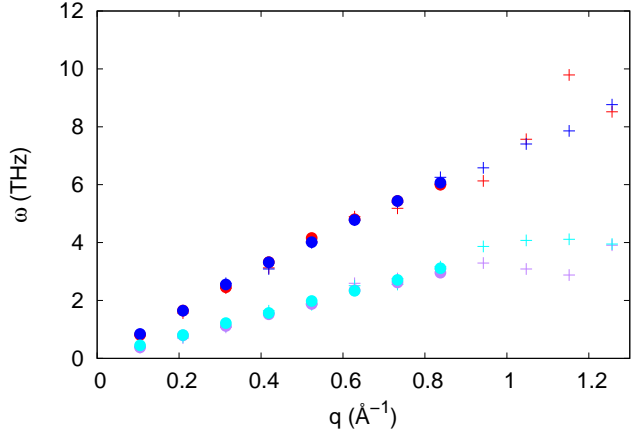}
 \caption{Acoustic dispersion obtained as the maximum of the dynamical structure factor for different $q$ (crosses) and from the "pseudo-experimental" method (circles) for the Amorphous sample (longitudinal in red, transverse in purple) and CinA30 sample (longitudinal in blue, transverse in cyan).}\label{dispersionA}
\end{figure}

In order to get quantitative information from our results, we have calculated the frequency and linewidths of acoustic modes using two procedures: one is to do a statistical analysis of the theoretical data; the second one is to artificially create a pseudo-experimental spectrum that we can thus analyze using standard fitting procedures. For both methods, the acoustic dispersion is obtained from the maximum of the dynamical structure factor at each $q$ (Fig.~\ref{dispersionA}). As for the width, different calculation methods can also be used on the theoretical data~\cite{thomas_predicting_2010, beltukov_ioffe-regel_2013, buchenau_2014}. 

For the first method, we compute the width with the standard deviation formula, as: \begin{equation}\Delta\omega(q)\sim\left(\frac{\int_0^{\omega_{max}}(\omega-<\omega>_q)^2S(\omega) \, \mathrm d\omega}{\int_0^{\omega_{max}}S(\omega) \, \mathrm d\omega}\right)^{1/2}\label{eqdelta}\end{equation}

\begin{figure}
\includegraphics[scale=0.7]{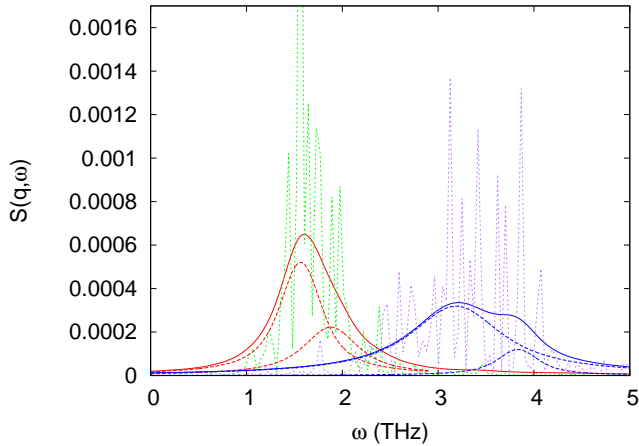}
\caption{Result of the convolution of the calculated $S(q,\omega)$ with an experimental energy resolution function (see text for details) at two different wavevectors: $q=2.09~\angstrom^{-1}$ (in red) and $q=4.18~\angstrom^{-1}$ (in blue). Notice that at high $q$ the convolution clearly shows an envelope of several modes when at low wavevector the dynamical structure factor is almost unimodal. The dashed lines show the peaks used to fit the spectrum.}
  \protect\label{conv}
\end{figure}

\begin{figure}
\includegraphics[scale=0.7]{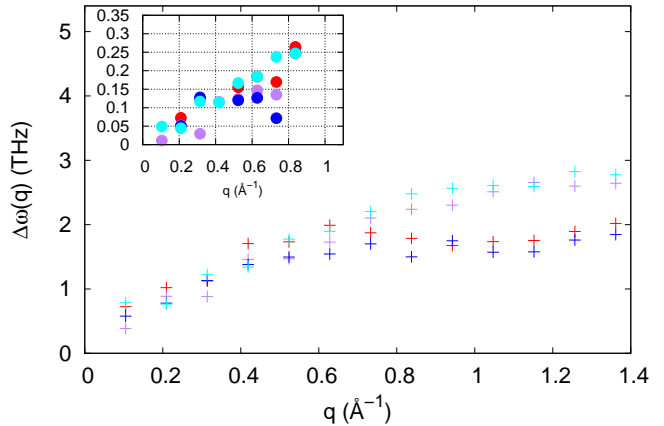}
\caption{Linewidth given by the standard deviation of the dynamic structure factor (formula~\ref{eqdelta}) for the amorphous sample (longitudinale in red, transverse in purple) and the CinA30 sample (longitudinale in blue, transverse in cyan). Inset: linewidths given by a fit of the dynamic structure factor with the lorentzian function.}
\protect\label{domega}
\end{figure}

\begin{figure}
\includegraphics[scale=0.7]{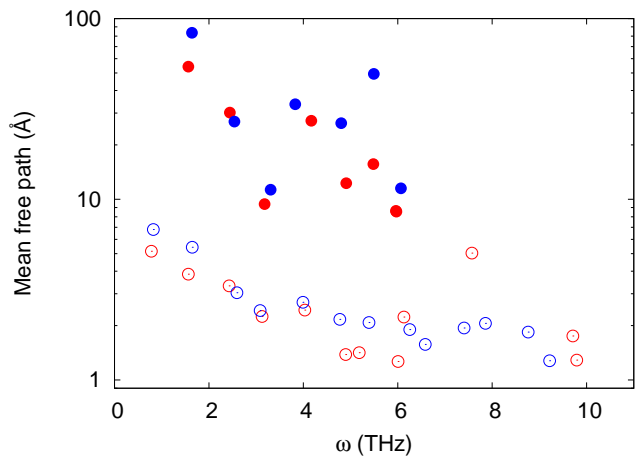}
\caption{Longitudinal modes MFPs for the amorphous (in red) and CinA30 (in blue) samples obtained from the linewidth of the dynamical structure factor with the two methods (see text): pseudo-experimental (full circles) and standard deviation (empty circles).}
  \protect\label{MFP_dsf}
\end{figure}

For the second method, in order to create a pseudo-experimental spectrum, we have convoluted our data with a typical energy resolution function with a linewidth of $\approx 1.4$~meV, as measured on the x-ray inelastic scattering beamline ID28 of the European Synchrotron Radiation Source (ESRF). The convolution gives rise to a nice single peak for low momentum transfers, while at high $q$ we clearly see an envelope of many modes (see Fig.~\ref{conv}). 
These pseudo-experimental data have then been fitted using  a sum of Lorentzian functions as a model for the inelastic excitation. The model is then convoluted with the energy resolution prior to be fitted to the data. We find that the most intense peak of the envelope at high $q$ is indeed the same mode as the single peak at low $q$, giving rise to the acoustic dispersion. 
The acoustic dispersions for longitudinal and transverse modes obtained by the two methods are in very good agreement (Fig.~\ref{dispersionA}). In Fig.~\ref{domega} we report the full width at half maximum (FWHM) of the most intense Lorentzian peak for the longitudinal and transverse modes for the amorphous and the partially crystalline samples. 
We can notice that the linewidths obtained through the "pseudo-experimental" procedure are approximately 10 times smaller than the ones obtained through the mean deviation formula. This is expected since in the former case the fit only takes into account a single peak and its width. On the other hand the standard deviation looks at the distribution for all frequencies, including high frequencies where significant values of the dynamical structure factor can be seen. The linewidth as obtained from the pseudo-experimental data is thus more meaningfull. Beside this, the two methods show the same evolution for the linewidths with higher values for the transverse than for the longitudinal modes above $q=0.7~\angstrom^{-1}$. Once the dynamical structure factor is fitted,  the width of the peak $\Delta\omega$ (called damping parameter in the DHO model) can be related to the lifetime ($\tau^*$) of the phonon at frequency $\omega$. If we set $\tau^*(\omega)\simeq\frac{1}{\Delta\omega}$, a mean free path (MFP) can be computed as \begin{equation}l^*(\omega)=v_g(\omega)\tau^*\label{eqdelta_MFP}\end{equation} where $v_g(\omega)$ is the group velocity at frequency $\omega$ as far as this velocity is well defined (Fig.~\ref{MFP_dsf}). In amorphous materials, eigenmodes have limited lifetime due to different contributions like scattering by the disordered organization of the particles. It should be underlined that the dynamical structure factor cannot be fitted correctly for all frequencies: the Ioffe-Regel criterion, defined as $l_{IR}=\lambda /2=\pi /q$ usually sets the limit between a propagative and diffusive transport of energy and heat~\cite{allen_diffusons_1999}. This sets a critical $q$ beyond which the phonon's wavevector is ill-defined and the phonon cannot be seen as a plane wave. Moreover, $v_g$ is ill-defined for $q\ge0.7~\angstrom^{-1}$, where the modes become non dispersive, thus the MFP cannot be calculated with Eq.~\ref{eqdelta_MFP}, but other methods based on diffusivity are required~\cite{larkin_2014}. Interestingly, it is shown from the dispersion relation (Fig.~\ref{dispersionA}), that the frequencies corresponding to this wavevector correspond as well to the frequencies that bound the Boson peak seen in Fig.~\ref{vdos}, $2.5~THz$ being the corresponding frequency of transverse waves and $6~THz$ being the frequency of longitudinal waves. In this frequency range, the longitudinal MFP, which is still well defined, decays monotonously (Fig.~\ref{MFP_dsf}), in agreement with similar calculations realized at lower frequencies by Larkin \textit{et al.} in Ref.~\cite{larkin_2014}. This decay is expected since the waves should have MFPs of the order of the size of the system at high wavelengths and be more and more scattered as the wavelengths tend toward the size of the local disorder where the wave is strongly scattered.

\subsection{Propagation of a Wave Packet}

An alternative way to measure MFPs is to look at the real space propagation of a wave packet. To do this, we impose a wave packet to a central layer ($4~\angstrom$ wide) of a sample during a MD run as done in reference~\cite{beltukov_ioffe-regel_2013}: the wave packet is created by imposing a displacement to the particles: \begin{equation}\vec{u}(t)=A\exp(\frac{-t^2}{2\tau^2})\sin(\omega t)\vec{e}\end{equation}
with $A=49.10^{-2}/\omega$ in order to limit the kinetic energy, with a maximum displacement of $49.10^{-3}~\angstrom$ for $\omega~=~1~THz$, ensuring that anharmonic effects are negligible. The parameter $\tau$ controls the decay of the amplitude of the wave with time and was set to 1~ps. This value is one tenth of the total simulation time of 10 ps. Longitudinal and transverse wave packets can be considered separately depending on the relative orientation of $\vec{e}$ and $\vec{e}_x$ ($\vec{e}=\vec{e}_x$ for longitudinal, and $\vec{e}.\vec{e}_x=0$ for transverse waves).

\begin{figure*}
  \centering
 \includegraphics[width=.45\linewidth]{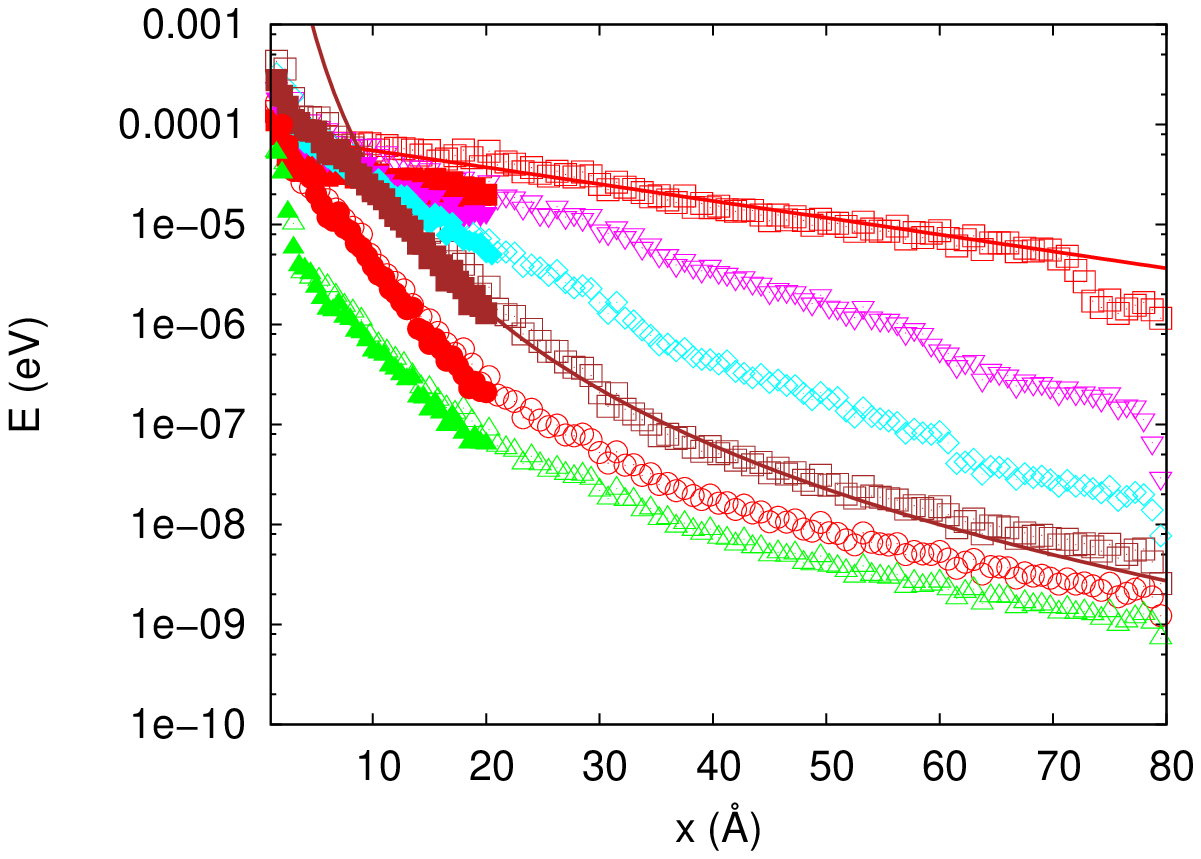} \qquad
  \includegraphics[width=.45\linewidth]{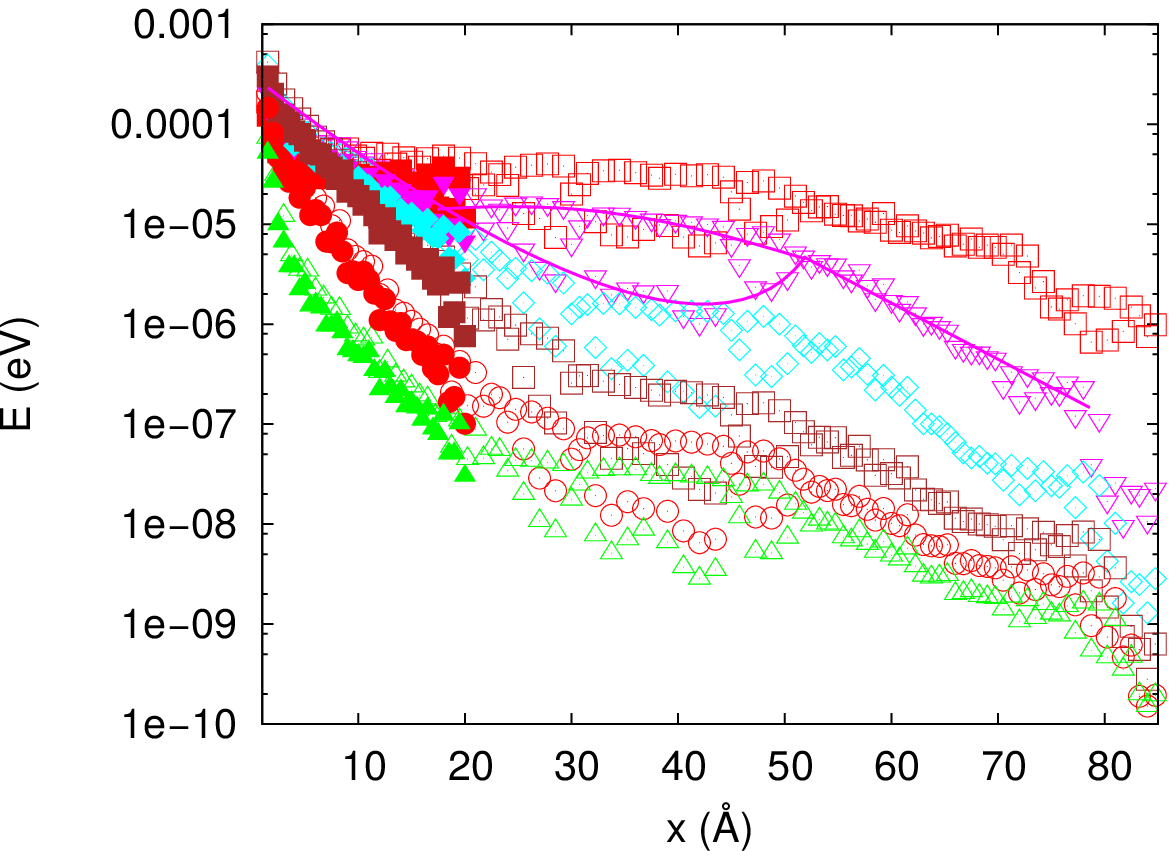}
  \caption{Linear-log representation of the kinetic energy envelopes resulting from the propagation of longitudinal waves in the amorphous (on the left) and CinA30 (on the right) sample. The filled symbols (limited to $20~\angstrom$) are the results obtained on the small systems described by the dynamical matrix and the empty symbols are the results obtained on the long systems described by the SW potential. On the figure only some of the frequencies have been reported: 1~THz (red square), 5THz (purple triangle), 7.5~THz (cyan diamond), 10~THz (brown square), 12.5~THz (red circle) and 15~THz (green delta). On the left figure, the red line represents an exponential fit from which the MFPs are obtained, here at frequencies 1~THz. The brown line is a power law defined as $a/x^{3.2}$ that fits the propagation of a wave of frequency 10~THz. The purple line on the right figure is a guide for the eye.}\protect\label{Elinearlog}
\end{figure*}

\begin{figure}
\includegraphics[scale=0.7]{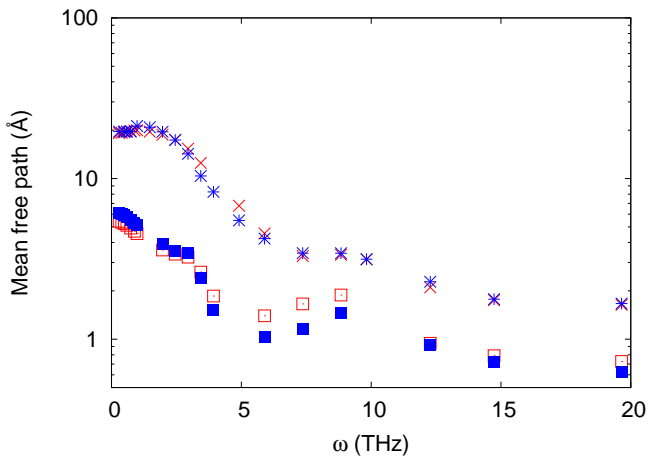}
\caption{MFP of transverse (squares) and longitudinal (crosses) plane waves as a function of the frequency for the amorphous (in red) and CinA30 (in blue) samples obtained through the propagation of wave packets.}
  \protect\label{mfp}
\end{figure}

From the propagation of the displacement and thus energy along the x-axis of the sample, MFPs $l(\omega)$ as a function of the frequency were obtained. For that, we computed first the total kinetic energy evolution with respect to the distance from the wave imposition layer ($x=0$): \begin{equation}E(x,t)=\sum_{i}^NE_i^K(t)~\delta(|x_i-x|)\end{equation} where $E_i^K$ it the kinetic energy of the ith particle. Plotting $E(x,t)$ for all the time steps during the propagation of the wave, we obtain a figure representing the repartition of the kinetic energy with time along the sample, Fig.~\ref{mfp}. 
The initial systems are $60~\angstrom$ large with periodic boundary conditions. Once a wave reaches the border of a system (after $30~\angstrom$), it comes back the other way and its energy superimposes with the energy of the leaving part of the wave. The simulations lasting for 10 ps, the longitudinal wave has time to travel  $77~\angstrom$ during the time of the simulation if we refer to the sound speed found previously. This distance is bigger than the size of the system. It limits the good fitting of the energy progression in the systems to approximately $30~\angstrom$, especially at low frequencies where the MFPs are expected to be of the order of the system size. To overcome this limitation we looked at the propagation of waves in longer systems made out of the repetition in the x direction of our original sample. In these longer systems, we chose to describe the interaction between particles using the SW potential that allows to take into account anharmonic effects. The resulting kinetic energy envelopes for both the small systems described by the dynamical matrix and the big systems described by the SW potential are shown in Fig.~\ref{Elinearlog}, confirming that anharmonic effects are here negligible.

\begin{figure}
\includegraphics[scale=0.7]{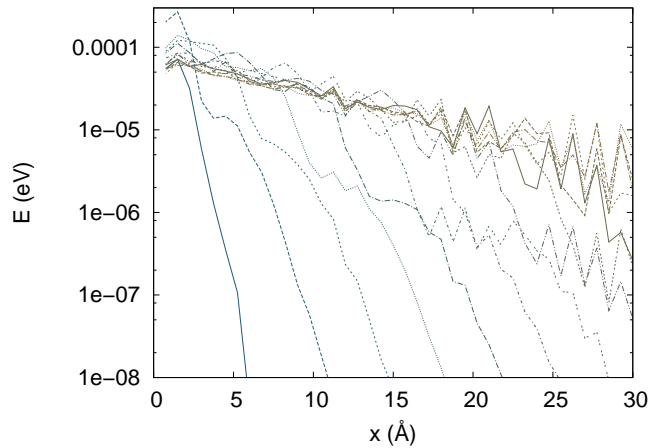}
\caption{Kinetic energy repartition for different time steps (t=n*800~fs) as a wave of frequency 5~THz propagates in the CinA30 sample.}
  \protect\label{energierep}
\end{figure}

While the energy repartition in the amorphous sample shows a continuous behavior, an original feature is observed in Fig.~\ref{Elinearlog} for CinA30. Indeed, from 15 to 50~$\angstrom$, the energy envelopes show a split in the repartition of energy and this at all frequencies.  In order to understand this split, we looked at the shape of the kinetic energy profile against time during the propagation of the wave, shown in Fig.~\ref{energierep}. We see here that, during the propagation, a shoulder appears in the spatial dependence of the kinetic energy, giving rise to two local maxima in the spatial distribution of kinetic energy and consequently to a split in the energy envelope. To understand the origin of this phenomenon, it is important to look at the geometry of the system. The CinA30 sample is 60~$\angstrom$ long and contains a spherical inclusion in its center of radius 24.92~$\angstrom$. As said, this configuration has been repeated several times in the x direction. Thus, in the direction of propagation of the wave, interfaces between crystal and amorphous material exist at x=5.08~$\angstrom$, 54.92~$\angstrom$ and for the following cell at 65.08~$\angstrom$ and 114.92~$\angstrom$. The interfaces are not plane surfaces since the inclusions are spherical and thus these positions are the positions of the beginning and the end of the crystalline inclusions. The splitting we observe occurs exactly at the coexistence of the asperity and of the random matrix in the direction of propagation. It means that the presence of the interface induces a difference of speed between the high and low energy propagation, the former being slowed down in a stronger manner. This is the signature of dispersion, that is the existence of different velocities for different energy components of the wave. This could be schematically understood as a plane longitudinal wave that breaks up at the interface into several components with different velocities similarly to what has been  reported in ref.~\cite{schelling2004} in the case of a traveling wave packet at the interface between crystalline grains. Moreover, the inclusion acts as a low pass filter forcing the high energies (more than 1/10th of the maximum kinetic energy) to pass around and not through it. It means that not only the frequency, but also the initial conditions (amplitude of the wave packet) affect the propagation of the wave in the composite system. It is worth noticing that the induced heterogeneity of the propagation is not reflected in the average MFP.

Let's now look at the resulting apparent MFP. At low frequencies, both samples show an exponential decay of the kinetic energy (straight red fit in Fig.~\ref{Elinearlog}). At higher frequencies and large distances, we can observe a transition to a power law behavior, for the longitudinal wave packet only ($\vec{e}=\vec{e}_x$). The characteristic frequency that marks the separation between these two behaviors is 2.5~THz. This change of trend can be interpreted as the crossing from simple to multiple scattering~\cite{van_tiggelen_multiple_1992}. However, multiple scattering behavior is usually fitted by an Ohm's like law ($l/x$)~\cite{van_tiggelen_multiple_1992} and we had to use a power law ($l^{\alpha}/x^{\alpha}$ with $\alpha\approx 3$ for $\omega \ge 2.5 \,THz$ see Table~\ref{Tab1}) to match the energy repartition of our data. Concerning the low frequency behavior of the transverse wave packets (with $\vec{e}.\vec{e}_x=0$), it can be fitted with a decreasing exponential function of the form $B\times\exp{(-x/l)}$ known as the Beer-Lambert law, from which $l$, the MFP of the wave, can be found. For the CinA30, since the energy envelope is not homogeneous, we chose to fit the average of the curve in order to obtain the MFPs. These $l(\omega)$ can be compared to those found through the fitting of the dynamical structure factor, $l^*(\omega)$ for $q<0.7~\angstrom^{-1}$. The behavior is similar, although the mean-free path obtained from the standard deviation formula of the structure factor is underestimated. For the longitudinal waves, an almost constant MFP can be observed in all cases at low frequency before a decrease starting at 2.5~THz with a minimum at 7.5~THz, followed by a regain around 9~THz already observed in Ref.~\cite{he_donadio_galli_2011}. The longitudinal MFP obtained from the wave-packet propagation decays from $20~\angstrom$ to $2~\angstrom$. The low frequency value is probably limited by the finite box size: indeed it is much smaller than the one obtained from the linewidth which is in agreement with Ref.~\cite{he_donadio_galli_2011}. The MFP for the transverse waves is always smaller than the one for the longitudinal waves, starting at $6~\angstrom$ at small frequencies down to $0.6~\angstrom$ at larger frequencies. It follows a similar trend with a saturation followed by a regain at 9~THz. This regain is unexpected since the transverse waves have been shown to disappear after the boson peak~\cite{Parshin2014} but could be explained only by a coupling between longitudinal and transverse modes during the dynamical propagation of the wavepacket. The largest discrepancy between amorphous and CinA30 sample occurs for transverse waves in the 2 to 12~THz range.
From these sets of data however it is not possible to definitively assess which sample between the amorphous and CinA30 is more resistive to the propagation of waves and energy, because the effect of the inclusion on the average mean-free path appears to be very small.

\begin{table}
\tiny
\resizebox{8.5cm}{!} {

\begin{tabular}{c|c|c|c|c|c|c|c}
\hline
$\omega$ (THz)& $  0.3 $ to $ 1 $& $1.5$ & $2 $&$2.5 $&$ 3 $&$ 3.5 $&$ 5$ \\
\hline
$\alpha$&-&1.9&2&2.2&2.5&2.8&3\\
\hline
\hline
 $\omega$ (THz)&$ 6$&$ 7$&$ 9$&$ 10$&$ 12$&$ 15$&$ 20$  \\
\hline
$\alpha$& 3&3&3.1&3.2&3.3&3.2&2.9\\
\hline

\end{tabular}
}
\caption{\label{Tab1} Exponent $\alpha$ of the power-law fit of the large distance energy envelope, as a function of the frequency $\omega$, for longitudinal wave packets.}
\end{table}

While MFP obtained from the wave propagation and from the Lorentzian fits are similar, the detailed study of the waves propagation show evidence of spatial heterogeneities. In the amorphous sample, we have observed a transition from simple scattering-like to multiple scattering-like in the attenuation of energy of a longitudinal wave-packet, at a scale comparable to the measured MFP. In the presence of the inclusion (whose size is larger than the MFP of the amorphous matrix), the crystallite acts as a low pass filter for the kinetic energy, adding fluctuations to the spatial envelope of the energy at larger scale than the MFP. These heterogeneities cannot be catched by the MFP, but still they must  play an important role in thermal transport, thus raising the question of the correct modeling of thermal transport in composite systems.

\section{Conclusions and Discussion}

In this paper, we studied in detail the role of nanometric crystalline inclusions on the vibrational properties of amorphous samples (a numerical model of Si  sample). We find that the VDOS of a sample with inclusions is approximately described by the sum of the VDOS of the amorphous and of the crystalline samples, although subtle differences still persist. However, the coefficients of the linear combination are not proportional to the volumic fraction of crystal in the sample, indicating that this is not the right parameter. The main effect of the inclusions on the density of states is to lower the number of resonant modes at frequencies close to 12.5~THz, a frequency where the density of vibrational modes is very low in crystalline samples.
The average MFP measured here in different ways appears to be insensitive to this peculiar frequency, and does not show a strong dependence on the inclusion. 
One visible effect of the presence of the inclusion is a decrease of the MFP of  transverse waves for frequencies between 2.5 and 10~THz. These frequencies are close to the frequencies bounding the Boson Peak in the amorphous sample, that were related in this paper to the frequencies at which transverse (2.5 THz) and longitudinal (7.5 THz) modes are significantly scattered by the amorphous structure. 
Interestingly, from a close inspection of the MFP as obtained looking to the propagation of wave packets, we find that longitudinal wavepackets are already multiply scattered at large distance for frequencies smaller than 7.5 THz. However the related longitudinal mean-free path is far larger than the transverse one, and no effect is visible in the dynamical structure factor.
The two frequencies mentioned above appear in the respective dispersion relations for the same value of the wavevector $q=0.7~\angstrom^{-1}$ that is a wavelength $\lambda=2\pi/q=9~\angstrom$. This observation rises the question of the existence of a characteristic length for waves scattering in amorphous samples~\cite{tanguy_continuum_2002}.
This mesoscopic length is larger than the typical length scale for plastic rearangement (5 to 6~\angstrom~\cite{fusco_role_2010}) but of the same order of the mean-free paths and of the distance above which the Beer-Lambert law for acoustic attenuation breaks down to a Ohm's like law. The fact that the transition from simple to multiple scattering is seen only in longitudinal waves could be due to the fact that the transverse mean-free path is always smaller that this characteristic distance. 
On the other hand, this length cannot be linked easily to any structural feature in our amorphous samples: larger than the average first-neighbour distance, it could be related to long range correlations due to disorder~\cite{tanguy_continuum_2002} and elastic heterogeneities~\cite{marruzzo_schirmacher_2013}. However, further work is still needed to understand if it can be related to some characteristic local order existing in amorphous silicon.

The main effect of the crystalline inclusion is to act as a low-pass filter for the kinetic energy. At the interface between amorphous and crystalline parts, the high kinetic energies are slowed down more strongly than the low kinetic energies. This results in an amplified heterogeneous waves propagation, and in enhanced fluctuations in the energy envelope, thus raising the question of the mean-free path definition in this case, as well as its role on thermal transport~\cite{kreuzer_1981}. A deeper understanding of the microscopic origin of this unexpected kinetic effect would allow to design nanostructured samples with very high acoustic attenuation.

More work is needed for investigating the effect on thermal transport of size and shape of the crystal inclusion as well as the possible interaction when there are more than one inclusion. Due to the relativily small size of the systems that can be studied using MD, other methods are advisable as it has been done by Zhang \textit{et al.} in Ref.~\cite{zhang_best_2014}.
\\

\begin{acknowledgments}

Authors thank the financial support of ARC Energie (r\'egion Rh\^{o}ne-Alpes) and Labex IMUST. Stimulating discusions with J.-J. Blandin and S. Gravier (SIMAP), T. Albaret, S. Merabia and S. Pailh\`es (ILM) are gratefully acknowledged.

\end{acknowledgments}

\bibliographystyle{unsrt}
\bibliography{bibliographie}

\end{document}